\documentclass[preprint,12pt]{elsarticle}
\makeatletter
\def\ps@pprintTitle{%
	\let\@oddhead\@empty
	\let\@evenhead\@empty
	\def\@oddfoot{}%
	\let\@evenfoot\@oddfoot}
\makeatother
\usepackage{latexsym,amssymb,amsmath,epsfig,amsthm}
\usepackage{floatrow} % figure won't float around ...
\usepackage{parskip} % avoid horizontal spacing for the first sentence in a paragraph
\usepackage{amsthm}
\usepackage{upgreek}
\usepackage[super]{nth}
\usepackage[makeroom]{cancel}
\usepackage{hyperref}
\usepackage{array,booktabs}
\usepackage{bm}
\usepackage{siunitx}
\usepackage{tikz}
\usepackage{algorithm}
\usepackage{algpseudocode}
\usepackage{url}
\usepackage[normalem]{ulem}
\graphicspath{{figs/}}
\numberwithin{equation}{section}
\usepackage{mathtools}
\usepackage[symbols,nogroupskip,record]{glossaries-extra}
\usepackage{longtable}
\glsxtrnewsymbol[description={Electrolyte salt concentration [mol/m$^3$]}]{$u(x,t)$}{\ensuremath{u(x,t)}}
\glsxtrnewsymbol[description={Solid-phase surface concentration [mol/m$^3$]}]{$c^*(x,t)$}{\ensuremath{c^*(x,t)}}
\glsxtrnewsymbol[description={Electrolyte diffusivity [m$^2$/s]}]{$D_i$}{\ensuremath{D_i}}
\glsxtrnewsymbol[description={Solid-phase diffusivity [m$^2$/s]}]{$D^s_i$}{\ensuremath{D^s_i}}
\glsxtrnewsymbol[description={Reaction rate [m$^{2.5}$/(mol$^{0.5}$s)]}]{$k_i$}{\ensuremath{k_i}}
\glsxtrnewsymbol[description={Thickness [m]}]{$l_i$}{\ensuremath{l_i}}
\glsxtrnewsymbol[description={Particle radius [m]}]{$R_{p,i}$}{\ensuremath{R_{p,i}}}
\glsxtrnewsymbol[description={Density [kg/m$^3$]}]{rho}{\ensuremath{\rho_i}}
\glsxtrnewsymbol[description={Specific heat [J/(kg$\cdot$K) ]}]{$C_p$}{\ensuremath{C_{p,i}}}
\glsxtrnewsymbol[description={Thermal conductivity [W/(m$\cdot$K)]}]{lambda}{\ensuremath{\lambda_i}}
\glsxtrnewsymbol[description={Solid-phase conductivity [S/m]}]{sigma}{\ensuremath{\sigma_i}}
\glsxtrnewsymbol[description={Porosity [-]}]{epsilon}{\ensuremath{\epsilon_i}}
\glsxtrnewsymbol[description={Particle surface area to volume [m$^2$/m$^3$]}]{a}{\ensuremath{a_i}}
\glsxtrnewsymbol[description={Reaction constant activation energy [J/mol]}]{Ek}{\ensuremath{E^{k_i}_a}}
\glsxtrnewsymbol[description={Solid-phase diffusion activation energy [J/mol]}]{Ek}{\ensuremath{E^{D^s_i}_a}}
\glsxtrnewsymbol[description={Bruggeman's coefficient [-]}]{brugg}{brugg}
\glsxtrnewsymbol[description={Faraday's constant [C/mol]}]{F}{\ensuremath{F}}
\glsxtrnewsymbol[description={Universal gas constant [J/(mol$\cdot$K)]}]{R}{\ensuremath{R}}
\glsxtrnewsymbol[description={Porosity [-]}]{epsilon}{\ensuremath{\epsilon_i}}
\glsxtrnewsymbol[description={Transference number [-]}]{t}{\ensuremath{t_+}}
\glsxtrnewsymbol[description={Filler fraction [-]}]{epsilonf}{\ensuremath{\epsilon_{f,i}}}
\usetikzlibrary{shapes.geometric, arrows}
\tikzstyle{startstop} = [rectangle, rounded corners, minimum width=3cm, minimum height=1cm,text centered, text width=3.5cm,draw=black]
\tikzstyle{process} = [rectangle, minimum width=3cm, minimum height=1cm, text centered,text width=3.5cm, draw=black]
\tikzstyle{io} = [trapezium, trapezium left angle=70, trapezium right angle=110, minimum width=3cm, minimum height=1cm, text centered,text width=3.5cm, draw=black]

\tikzstyle{arrow} = [thick,->,>=stealth]

\newcommand{\del}{\ensuremath{\partial}}
\newcommand{\eff}{\ensuremath{\text{eff}}}
\newcommand{\bigoh}{\ensuremath{\mathcal{O}}}
\newcommand{\mbf}{\ensuremath{\mathbf}}

\newcommand*{\DisplayEq}[1]{%
	\ensuremath{\displaystyle #1}%
}
\newcommand*{\DefEq}[2]{%
	\begin{tabular}{@{}l@{}}%
		#1\\
		\quad\DisplayEq{#2}%
	\end{tabular}%
}

\begin{document}

\begin{frontmatter}



\title{A fast solver for the pseudo-two-dimensional model of lithium-ion batteries}

\author[1]{Rachel Han}
\ead{hanrach@math.ubc.ca}

\author[1]{Colin Macdonald}
\ead{cbm@math.ubc.ca}

\author[1]{Brian Wetton}
\ead{wetton@math.ubc.ca}

\affiliation[1]{organization={University of British Columbia}, addressline={6356 Agricultural Road},
	postcode={V6T 1Z2}, city={Vancouver}, country={Canada}}

%\affiliation[2]{organization={Sayahna Foundation}, addressline={JWRA 34, Jagathy},
%	city={Trivandrum}
%	postcode={695014},
%	country={India}}
%\affiliation[3]{organization={Sayahna Foundation}, addressline={JWRA 34, Jagathy},
%	city={Trivandrum}
%	postcode={695014},
%	country={India}}
\begin{abstract}

The pseudo-two-dimensional (P2D) model is a complex mathematical model that can capture the electrochemical processes in  Li-ion batteries. However, the model also brings a heavy computational burden. Many simplifications to the model have been introduced in the literature to reduce the complexity. We present a method for fast computation of the P2D model which can be used when simplifications are not accurate enough. By rearranging the calculations, we reduce the complexity of the linear algebra problem. We also employ automatic differentiation, using an open source package \texttt{JAX} for robustness, while also allowing easy implementation of changes to coefficient expressions. The method alleviates the computational bottleneck in P2D models without compromising accuracy.

\end{abstract}


\end{frontmatter}


\section{Introduction}\label{sec:intro}
Lithium-ion (Li-ion) batteries are essential in modern energy storage and widely used in devices from portable electronics to large electric vehicles. Li-ion batteries have high energy efficiency and high power density compared to their predecessors such as lead-acid and zinc-carbon batteries \cite{Nitta2015}. They garnered great research interest since their commercial inception in the 1980s \cite{heelan2016current}.

However, the battery needs to be monitored for control and optimization purposes, especially for complex systems. This is due to the nonlinear internal behaviour in the battery that is difficult to capture with external observations. The battery management system (BMS) plays a key part in this purpose. Its functions include predicting the state of charge (SOC) (the level of charge given the battery capacity), the state of health (SOH) (a qualitative measure of how much the battery has degraded from its original specifications) and the remaining useful life (RUL) \cite{Pattipati2008}. These characteristics cannot be directly measured and can only be inferred by models.

The battery management system battery models can largely be classified into two categories: equivalent circuit models and electrochemical models. In more recent literature, data-driven methods such as deep learning have been developed that try to predict internal battery behaviour, as done in \cite{liu2016}, \cite{Severson2019} and \cite{Ng2020}.
We focus on a particular electrochemical model called the pseudo-two-dimensional (P2D) model, also known as the Doyle-Fuller-Newman model \cite{Doyle_1993}. It is a full physics based model described by partial differential algebraic equations. This complex model allows for a more accurate description of the battery but also brings a heavy computational burden on a practical BMS. In the literature, there are many electrochemical models with reduced complexity such as the single particle model \cite{guo2010}, \cite{prada2012} and approximations of solid-phase diffusion \cite{ram2010}, \cite{han2015}. The simplified models are used in applications such as control, on-line monitoring, optimization, parameter estimation and age prediction \cite{jokar2016review}. In this work, we present a method for fast computation of the P2D model which can be used when the simplifications are not sufficiently accurate.

The nonlinear partial differential algebraic equations (PDAE) that describe the P2D model are often implemented by Finite Difference Methods (FDM), Finite Volume Methods (FVM) and Finite Elements Methods (FEM) \cite{Ramadesigan2012}. Commercial software such as COMSOL Inc. Multiphysics software \cite{COMSOL} has FDM and FEM implementations for the P2D model. \texttt{LIONSIMBA} \cite{torchio2016} uses FVM to implement the P2D model in MATLAB with reductions made in the solid-phase diffusion. It is typical in literature (\cite{botte2000mathematical}, \cite{torchio2016}) to use black box differential algebraic solvers, such as the \texttt{DASSL} solver \cite{petzold1982description} and \texttt{SUNDIALS} \cite{hindmarsh2005sundials} for time stepping.

In this paper, we develop a fast solver for the thermal P2D model taken from \cite{torchio2016} using FDM, backward Euler and Newton's method in Python. Additionally, we also employ automatic differentiation, using an open-source package \texttt{JAX} \cite{jax2018github}. This framework allows terms in the model to be changed easily, such as the diffusive and conductivity expressions. The method alleviates the computational bottleneck inherent in P2D models, without compromising accuracy. %This method could be further applied to other reduced models under certain assumptions.

\section{Thermal P2D model equations}\label{sec:p2d_pdae}

The pseudo-two-dimensional (P2D) model is an electrochemical model that captures the kinetics, transport processes and thermodynamics of a lithium-ion battery \cite{Doyle_1993}. The name pseudo-2D comes from the interesting simplification of the battery geometry. In the $x$ dimension, the processes in electrolyte phase across the battery are described. The second dimension, $r$, comes from the following assumption: in the porous electrodes, the host intercalation materials such as lithium cobalt oxide and graphite are represented by spherical particles at each channel location in $x$, where $r$ is in the direction normal to the surface of each particle. Moreover, $x$ is of macroscopic scale whereas $r$ is microscopic. P2D is not a strict two-dimensional model but rather a pseudo-2D model because the two dimensions with different scales are coupled. Figure \ref{fig:p2d} illustrates the simplified geometry of the P2D model.

\begin{figure}
	\centering
	\includegraphics[scale=0.5]{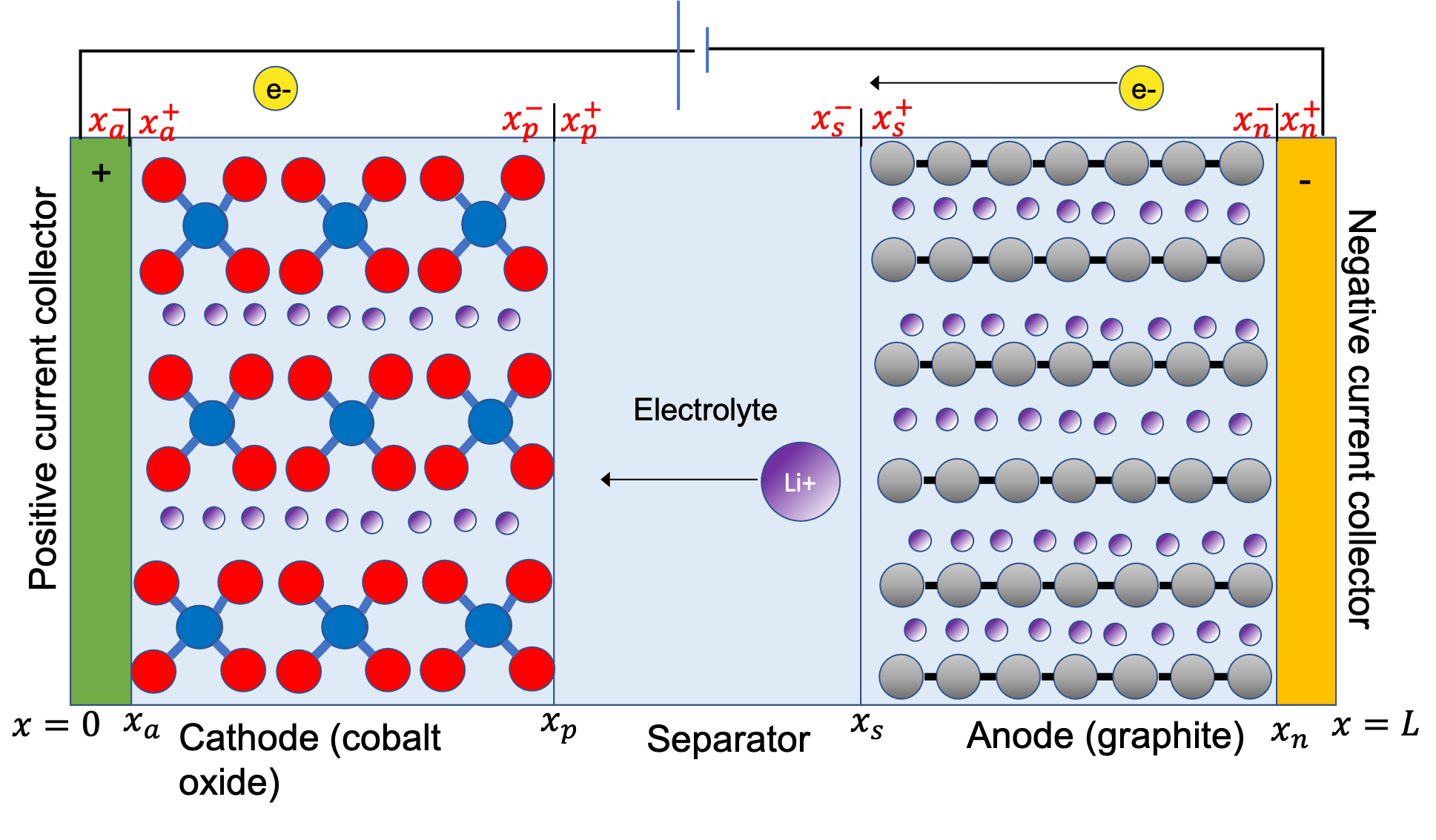}
	\caption{A representation of chemical properties of lithium-ion battery with lithium cobalt oxide and graphite. The purple spheres represent Lithium (Li) that are intercalated in cathode and anode active materials.}
	\label{fig:battery_chem}
\end{figure}

\begin{figure}
	\includegraphics[scale=0.45]{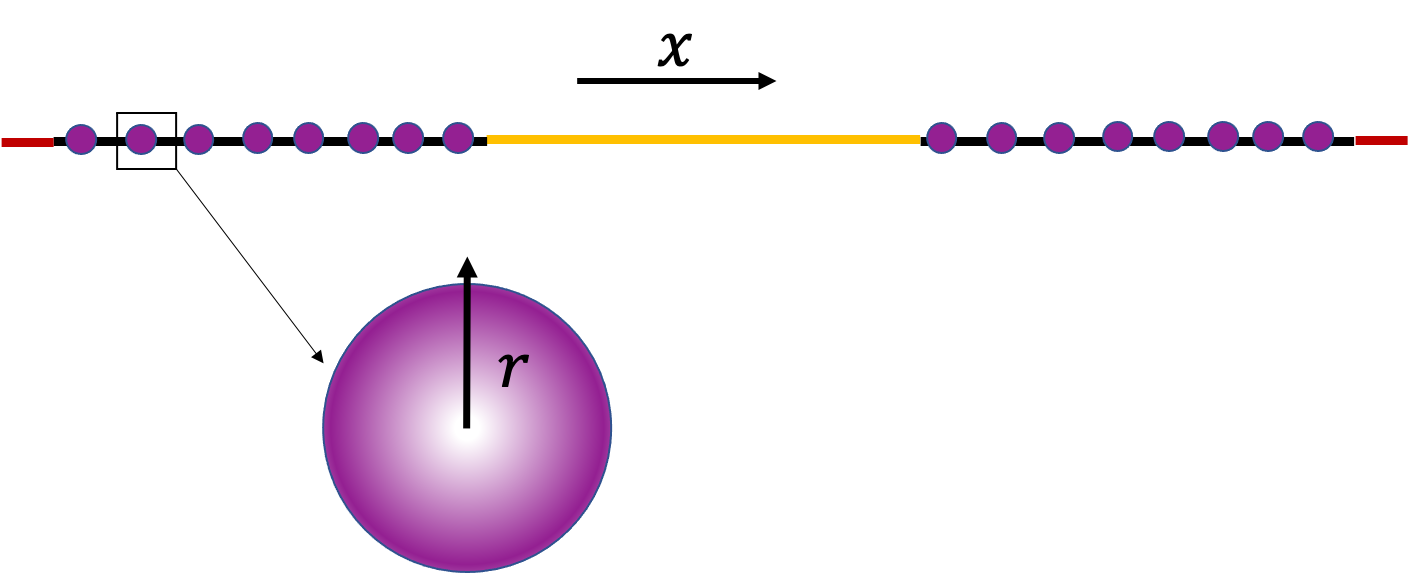}
	\caption{The P2D model abstraction of Figure~\ref{fig:battery_chem}}
	\label{fig:p2d}
\end{figure}

The P2D model is described by the following quantities: lithium-ion concentrations, potentials, ionic flux across the interface of the spherical particle, current density and temperature. These quantities describe the transport of lithium ion species, conservation of charge, kinetics of charge transfer and thermal behaviour of the battery. The lithium-ion concentrations are denoted by $u(x,t)$ in the electrolyte (liquid phase) and intercalated lithium concentration $c(x,r,t)$ in the solid particle (solid phase). The electric potentials in the liquid and solid phase are $\Phi(x,t)$ and $\Psi(x,t)$ respectively. The interfacial ionic flux is denoted by $j(x,t)$, $I$ is the current density and $T(x,t)$ is the temperature. We take the model form and coefficients from \cite{torchio2016}. The governing equations form a PDAE system, where the system is described by partial differential equations and algebraic constraints. We present the form of the equations in detail below, so that the reader can see how we exploit the structure of the nonlinear PDAE to obtain the fast solver.

Throughout the paper, we will use the notation $i \in \{a, p, s, n, z \}$ in subscript to denote the battery section: the positive current collector, cathode, separator, anode and the negative current collector respectively. The ends of the battery are at $x=0$ and $x=L$, where $L$ is the length of the battery. Figure \ref{fig:battery_chem} summarizes the domain. The subscript ``$\eff$'' is used to denote the effective coefficients based on Bruggeman theory \cite{bruggeman1935berechnung}, which accounts the for varying conductivity or diffusivity due to different materials in composite materials. The model parameters, coefficients and symbols are taken from \cite{torchio2016} and also listed in \ref{sec:appen_eqn} and \ref{sec:appen_param}.

\subsection{Electrodes and Separator}
The intercalation of Li$^+$ in the solid particles drive the charge and discharge in Li-ion batteries. The intercalation is driven by diffusion of the lithium concentration,
\begin{subequations} \label{eq:solid_conc_pde}
	\begin{equation}
	\frac{\partial c}{\partial t} = \frac{1}{r^2}\dfrac{\partial}{\partial r}\left( r^2 D^s \frac{\partial c(r,t)}{\partial r} \right),
	\end{equation}
	\begin{gather}
	\frac{\partial c(r,t)}{\partial r} \bigg\rvert_{r=0} = 0, \qquad \frac{\partial c(r,t)}{\partial r} \bigg\rvert_{r=R_p} = -\frac{j(x,t)}{D^s_{\eff, i}(T(x,t))}.
	\end{gather}
\end{subequations}
Here, $D^s$ is a constant solid-phase diffusion coefficient and $D^s_{\eff, i}$ for $i \in \{p, n\}$ is the effective solid-phase diffusion coefficient which depends on temperature in $x$. The kinetics of charge transfer at the electrodes are described by the Butler-Volmer equation,
\begin{equation}\label{eq:p2d_j}
j(x,t) = 2k_{\text{eff}}\sqrt{u(x,t)(c^\text{max}_{s,i} - c^*(x,t))c^*(x,t)}\sinh\left[\dfrac{0.5F}{RT}\eta_i(x,t)\right]
\end{equation}
for $i \in \{p, n\}$, where $\eta_i(x,t)$ denotes the overpotential. The equation \eqref{eq:p2d_j} is an algebraic constraint in the PDAE, along with the equation for $\eta_i(x,t)$ in \eqref{eq:p2d_eta}. In the porous electrodes $i \in \{p, n\}$, the accumulation of Li$^+$ in the electrolyte phase is described in terms of diffusion of the ions and $j$:
\begin{subequations} \label{eq:p2d_ueq}
	\begin{equation}
	\epsilon_i \frac{\partial{u(x,t)}}{\partial t}  = \frac{\partial}{\partial x} [ \mathbf{D_{eff, i}} \frac{\partial u(x,t)}{\partial x} ] + a_i(1-t_+)j(x,t),
	\end{equation}
	with equal flux conditions enforced at the boundaries,
	\begin{gather}
	\frac{\partial u(x,t)}{\partial x} \bigg\rvert_{x = \hat{x}_0, \hat{x}_n} = 0, \\
	-\mathbf{D_{eff,p}} \frac{\partial u(x,t)}{\partial x} \bigg\rvert_{x = \hat{x}^-_p} = -\mathbf{D_{eff,s}} \frac{\partial u(x,t)}{\partial x} \bigg\rvert_{x = \hat{x}^+_p}, \\
	-\mathbf{D_{eff,s}} \frac{\partial u(x,t)}{\partial x} \bigg\rvert_{x = \hat{x}^-_s} = -\mathbf{D_{eff,n}} \frac{\partial u(x,t)}{\partial x} \bigg\rvert_{x = \hat{x}^+_s}.
	\end{gather}
\end{subequations}
The coefficient $\bm{D}_{\eff, i}$ is the effective diffusive coefficient of Li$^+$ in the electrolyte. The interface positions $\hat{x}_i$ are shown in Figure \ref{fig:battery_chem}. 

In the separator where there are no solid particles, we have $j=0$ which yields 
\begin{subequations}
	\begin{equation}
	\epsilon_i \frac{\del u(x,t)}{\del t} = \frac{\del}{\del x} \left[ D_{\eff, s} \frac{\del u(x,t)}{\del x} \right],
	\end{equation} 
	with boundary conditions
	\begin{gather}
	-\bm{D}_{\eff,p} \frac{\del u(x,t)}{\del x} \rvert_{x = \hat{x}^-_p} =-\bm{D}_{\eff,s} \frac{\del u(x,t)}{\del x} \rvert_{x = \hat{x}^+_p}, \\
	-\bm{D}_{\eff,s} \frac{\del u(x,t)}{\del x} \rvert_{x = \hat{x}^-_s} =-\bm{D}_{\eff,n} \frac{\del u(x,t)}{\del x} \rvert_{x = \hat{x}^+_s}.
	\end{gather}
\end{subequations} 

In the electrodes $i \in \{p, n\}$, the transport of Li$^+$ is described by $\Phi$, $T$, $u$ and $j$:
\begin{subequations}\label{eq:p2d_phieq}
	\begin{equation}
	a_iFj(x,t) = -\frac{\partial }{\partial x} \left[ \mathbf{\kappa_{eff, i}} \frac{\partial \Phi(x,t)}{\partial x} \right] +\frac{\partial }{\partial x} \left[ \mathbf{\kappa_{eff,i}} \Upsilon T(x,t) \frac{\partial \ln u(x,t)}{\partial x} \right],
	\end{equation}
	with boundary conditions
	\begin{gather}
	\frac{\partial \Phi(x,t)}{\partial x} \bigg\rvert_{x = \hat{x}_0} = 0,\\
	- \mathbf{\kappa_{eff,p}} \frac{\partial \Phi(x,t)}{\partial x} \bigg\rvert_{x = \hat{x}_p^-} = - \mathbf{\kappa_{eff,s}} \frac{\partial \Phi(x,t)}{\partial x} \bigg\rvert_{x = \hat{x}_p^+}, \\
	- \mathbf{\kappa_{eff,s}} \frac{\partial \Phi(x,t)}{\partial x} \bigg\rvert_{x = \hat{x}_s^-} = - \mathbf{\kappa_{eff,n}} \frac{\partial \Phi(x,t)}{\partial x} \bigg\rvert_{x = \hat{x}_s^+}, \\
	\Phi(x,t)\bigg\rvert_{x = \hat{x}_n} =0,
	\end{gather}
		where $\bm{\kappa}_{\eff,i}$ is the effective electrolyte phase conductivity.
\end{subequations}

Again, in the separator, the ionic flux $j$ is $0$. Therefore, the ionic transport in the separator is described by:
\begin{subequations}
	\begin{equation}
	0 = -\frac{\del}{\del x}\left[ \bm{\kappa}_{\eff,s} \frac{\del \Phi(x,t)}{\del x}\right] + \frac{\del}{\del x}\left[ \bm{\kappa}_{\eff,s} T(x,t) \Upsilon \frac{\del \ln u(x,t)}{\del x}\right]
	\end{equation}
	with boundary conditions
	\begin{gather}
	-\bm{\kappa}_{\eff,p} \frac{\del \Phi(x,t)}{\del x} \rvert_{x = \hat{x}^-_p} =-\bm{\kappa}_{\eff,s} \frac{\del \Phi(x,t)}{\del x} \rvert_{x = \hat{x}^+_p}, \\
	-\bm{\kappa}_{\eff,s} \frac{\del \Phi(x,t)}{\del x} \rvert_{x = \hat{x}^-_s} =-\bm{\kappa}_{\eff,n} \frac{\del \Phi(x,t)}{\del x} \rvert_{x = \hat{x}^+_s}.
	\end{gather}
\end{subequations}

The movement of electrons in the electrodes is governed by Ohm's law \cite{newman2004}, which involves the solid phase potential $\Psi$,
\begin{subequations}
	\begin{gather}
	\frac{\partial}{\partial x} \left[  \mathbf{\sigma_{eff,i}} \frac{\partial \Psi(x,t)}{\partial x} \right] = a_iFj(x,t), \\
	\mathbf{\sigma_{eff,i}} \frac{\partial \Psi(x,t)}{\partial x} \bigg\rvert_{x = \hat{x}_0, \hat{x}_n} = -I_{\text{app}}(t), \qquad
	\mathbf{\sigma_{eff,i}} \frac{\partial \Psi(x,t)}{\partial x} \bigg\rvert_{x = \hat{x}_p, \hat{x}_s} = 0.
	\end{gather}
\end{subequations}
For each electrode $i \in \{p, n\}$, $\sigma_{\eff, i}$ is the effective solid-phase conductivity and $I_{\text{app}}$ is the applied current.

The heat created in the battery is described by:
\begin{subequations}\label{eq:p2d_T}
	\begin{equation}
	\rho_iC_{p,i}\frac{\partial T(x,t)}{\partial t} = \frac{\partial}{\partial x}\left[ \lambda_i \frac{\partial T(x,t)}{\partial x}\right] + \bm{Q}_{\text{ohm}} + \bm{Q}_{\text{rxn}} + \bm{Q}_{\text{rev}},
	\end{equation}
	which includes different heat sources. As mentioned in \cite{torchio2016}, the ohmic generation rate $ \bm{Q}_{\text{ohm}}$ takes account of heat generated from the movement of Li$^+$, the reaction generation rate $\bm{Q}_{\text{rxn}}$ describes the heat resulting from the ionic flux and overpotentials, and finally the reversible generation rate $\bm{Q}_{\text{rev}}$ accounts for the change of entropy in the electrodes. Details of these expressions can be found in the \ref{sec:appen_eqn}. The thermal flux continuity interface conditions for temperature are enforced, 
	\begin{gather}
	-\lambda_z \frac{\partial T(x,t)}{\partial x} \bigg\rvert_{x = \hat{x}_0^-} = -\lambda_p \frac{\partial T(x,t)}{\partial x} \bigg\rvert_{x = \hat{x}_0^+}, \\
	-\lambda_n \frac{\partial T(x,t)}{\partial x} \bigg\rvert_{x = \hat{x}_n^-} = \lambda_z \frac{\partial T(x,t)}{\partial x} \bigg\rvert_{x = \hat{x}_n^+}
	\end{gather}
\end{subequations}
Similarly, the temperature effects in the separator is described but without $\bm{Q}_{\text{rxn}}$ and $\bm{Q}_{\text{rev}}$:
\begin{subequations}
	\begin{equation}
	\rho_i C_{p,i} \frac{\del T(x,t)}{\del t} = \frac{\del}{\del x} \left[ \lambda_i \frac{\del T(x,t)}{\del x} \right] + \bm{Q}_{ohm},
	\end{equation}
	\begin{gather}
	-\lambda_p \frac{\del T(x,t)}{\del x} \rvert_{x = \hat{x}^-_p} =-\lambda_s \frac{\del T(x,t)}{\del x} \rvert_{x = \hat{x}^+_p}, \\
	-\lambda_s \frac{\del T(x,t)}{\del x} \rvert_{x = \hat{x}^-_s} =-\lambda_n \frac{\del T(x,t)}{\del x} \rvert_{x = \hat{x}^+_s}.
	\end{gather}
\end{subequations}
Lastly, we give an equation for the surface overpotential $\eta(x,t)$ of the electrode $i \in \{p, n\}$:
\begin{equation}\label{eq:p2d_eta}
\eta_i(x,t) = \Psi(x,t) - \Phi(x,t) - \mathbf{U_i}(T(x,t), c(x,r,t))
\end{equation}
where $ \mathbf{U_i}$ is the open circuit potential which is fitted experimentally (see \ref{sec:appen_eqn}).

\subsection{Current Collectors}

The current collectors consist of conductive metal materials for transfer of electrons. We model the thermal effects in these regions:
\begin{subequations}
	\begin{align}
	\rho_i C_{p,i} \frac{\del T(x,t)}{\del t} = \frac{\del}{\del x} \left( \lambda_i \frac{\del T(x,t)}{\del x} \right) + \frac{I^2_{app}(t)}{\sigma_{\eff,i}},
	\end{align}
	with boundary conditions
	\begin{align}
	-\lambda_a \frac{\del T(x,t)}{\del x} \rvert_{x=0} &= h(T_{ref} - T(x,t)),\\
	-\lambda_z \frac{\del T(x,t)}{\del x} \rvert_{x=L} &= h( T(x,t) - T_{ref}),
	\end{align}
\end{subequations}
with heat exchange coefficient $h$. The boundary conditions represent Newton's law of cooling at the two ends of the battery, with ambient temperature $T_{ref}$.

\section{Implementation}
The system of equations in Section \ref{sec:p2d_pdae} is discretised in space by cell centred finite differences, and in time using backward Euler method for the time-dependent variables, $u(x,t)$, $c(x,r,t)$ and $T(x,t)$. The complete discretizations are given in Tables \ref{tab:discretisations} and \ref{tab:discretisations_separator}. %For brevity, we drop the subscript in the unknowns, $i\in\{a,p,s,n,z\}$ except when specifying the boundary conditions and domain-specific coefficients.
We describe the structure of the discrete nonlinear system below. 

\subsection{Solving the nonlinear system}\label{sec:p2d_fd}

We first describe how to solve the resulting nonlinear system from the PDAE without considering the specific structure of the problem. We will call this basic implementation as \textit{na\"{i}ve method}. After discretising in space, we obtain a set of differential algebraic equations (DAE). In particular, we obtain a semi-explicit index-1 DAE, where one can distinguish the differential and algebraic variables. Variables without time derivatives are called algebraic as in standard PDAE literature \cite{ascher1998}.  We denote the discrete versions of the electrochemical quantities $c(x,r,t), j(x,t), u(x,t), \Phi(x,t), \Psi(x,t), T(x,t)$ and $\eta(x,t)$ as 

\begin{align*}
\bm{c}  &= (c_p, c_n), \quad \bm{j} = (j_p, j_n), \quad \bm{u} = (u_p, u_s, u_n), \\
\bm{\Phi} &= (\Phi_p, \Psi, \Phi_n), \quad \bm{\Psi} = (\Psi_p, \Psi,_n), \\
\bm{T} &= (T_a, T_p, T_s, T_n, T_z), \quad \bm{\eta} = (\eta_p, \eta_n).
\end{align*}
We can group the differential variables in $x$ as $\hat{U} = [\bm{u}, \bm{T}]$ and algebraic variables as $\tilde{U} = [ \bm{\Phi}, \bm{\Psi}, \bm{j}, \bm{\eta}]$. The solid particle concentration $\bm{c}$ is also a differential variable in $r$. After discretising in time using backward Euler, we form the following set of nonlinear equations,
\begin{subequations}\label{eq:pdae_nonlinear}
	\begin{align}
	\bm{0} & = \bm{c}^k - \Delta t\bm{G}(\bm{c}^k,\hat{U}^k, \tilde{U}^k) - \bm{c}^{k-1}\\
	\bm{0} &= \hat{U}^k- \Delta t \bm{\hat{F}}(\bm{c}^k, \hat{U}^k, \tilde{U}^k) -\hat{U}^{k-1}\\
	\bm{0} &= \bm{\tilde{F}}(\bm{c}^k, \hat{U}^k, \tilde{U}^k),
	\end{align}
\end{subequations}
where $k$ denotes the time step. From \eqref{eq:pdae_nonlinear}, we form a larger multi-dimensional vector valued function $\bm{F}(U)$, where $U$ is the aggregate vector containing all the unknowns, 
\begin{equation}\label{eq:bigU}
\begin{split}
U &= 
[\begin{matrix} c_p & c_n & u_p & u_s & u_n& j_p& j_n& \eta_p & \eta_n \end{matrix} \\
&\qquad\qquad \begin{matrix}  \Phi_p & \Psi & \Phi_n & \Psi_p & \Psi_n & T_a & T_p & T_s & T_n & T_z \end{matrix}]^T.
\end{split}
\end{equation} 
Solving for \eqref{eq:pdae_nonlinear} is equivalent to finding the root of the function 
\begin{equation}\label{eq:F}
\bm{F}(U^k)=\bm{0}
\end{equation}
at every $k$th time step. We solve for the root of \eqref{eq:F} iteratively using Newton's iteration, 
\begin{equation}
U^{p+1} = U^p - [\mathcal{J}^{-1}(U^p)] \bm{F}(U^p)
\end{equation}
where $p$ denotes the $p$th iterate of the Newton's method and $\mathcal{J}$ is the Jacobian of $\bm{F}$ at $U^p$. Solving for the inverse of $\mathcal{J}(U^p)$ is computationally expensive, especially when we have a large system. Therefore, we use a linear solver to solve for $\delta^p = [\mathcal{J}^{-1}(U^p)] \bm{F}(U^p)$,
\begin{equation}\label{eq:linear_system}
\mathcal{J}(U)\delta^p = \bm{F}(U^p).
\end{equation}
We generate the iterates $U^{p+1}$ by 
\begin{equation}
U^{p+1} = U^p - \delta^p
\end{equation}
at every iteration until we converge to the root. In this way, we can approximate the solution to PDAE by repeatedly solving large linear systems.

To give an idea of the size of the nonlinear system in this application, we outline the number of unknowns. In each battery section $i \in \{ p, n\}$, there are $M_i$ points in $x$. For each point in $x$, there are $N_i$ points in $r$. Then, the unknowns representing the solid-phase concentration $\bm{c}$ has $M_p(N_p + 2) + M_n(N_n+2)$ points, where $N_i + 2$ accounts for the ghost points \cite{ghost} needed to impose boundary conditions at $r=0$ and $r=R_p$. Similarly, each of $u_p, u_s$ and $u_n$ has $M_p + 2$, $M_s + 2$ and $M_n + 2$ unknowns respectively, and the same applies for the unknowns $\bm{\Phi}$ and $\bm{T}$. For $\bm{j}$ and $\bm{\eta}$, there are $M_p$ and $M_n$ points respectively in the positive and negative electrode, since there are no boundary conditions for these variables. In total, there are $$ M_p(N_p+2) + M_n(N_n+2) + 3(M_p + M_s + M_n + 6) + (M_p + M_n + 4) + 2(M_p + M_n)$$ unknowns. For example, when we set the number of grid points in each battery section and for each solid particle to be $M=N=30$, we have $2456$ unknowns in the resulting system. The number grows quadratically because of the dimension $r$ of the particles. This emphasizes the need for fast and robust solver for the full P2D model. The main contribution of this work is a fast solver strategy to the linear systems using some of the special structure of the discretised P2D model. For speed and robustness, we also need to consider how to efficiently construct the Jacobians in Newton iterations.

\subsection{Automatic-differentiation for Jacobians}
The fastest method of computing the Jacobian is to manually differentiate the function components with respect to every element in $U^p$ a priori. Then, the cost of computation is just the function evaluation of the differentiated components. This can be done for relatively simple models. However, for the fully-coupled P2D model, manual differentiation is inconvenient due to complex equations and prone to errors in the implementation. As alternatives, one can use numerical, symbolic or automatic differentiation (AD). 
Numerical differentiation, which involves finite differences, is easy to implement and fast. With numerical differentiation, we must choose the step size $h$ which is especially difficult for systems with multiple scales. We refer the readers to \cite{burden2011numerical} for the details on how round-off and truncation errors depend on the step size $h$.

On the other hand, symbolic and automatic differentiation (AD) compute exact Jacobians. Software for AD records every operation done on the inputs of a function, then builds a graph that encodes all the operations \cite{baydin2017automatic}. Then the rules of differentiation are applied on every node of the graph. Symbolic differentiation produces a full expression for the derivative, then the numerical values are substituted. This approach, however, has a problem of expression swell causing slower performance, as pointed out by \cite{baydin2017automatic}. We use AD for better performance, which uses intermediate values of computations to calculate the derivatives on the fly. There are two ways to do this, the forward and reverse mode AD. For functions $f : \mathbb{R}^n \rightarrow \mathbb{R}^m$ where $ n > m$, the reverse mode is preferred because the Jacobian can be constructed in $m$ passes but also uses additional space. In contrast, the forward mode is preferred for when $m > n$ because it takes $n$ passes to build the Jacobian. When $m=n$, as is the case for our system, the forward mode is usually preferred due to the large memory usage with the reverse mode. However, in Section \ref{sec:jac}, we show how we can use the backward mode AD to efficiently build the Jacobians. For further details on how AD works, we refer the reader to \cite{baydin2017automatic}. Note that the use of symbolic/automatic differentiation allows changes in terms in the equations (effective diffusivity expressions, for example) with no user effort needed to modify the Jacobian. 

A particular AD software that is suitable for the Python environment is a library named \texttt{JAX} \cite{jax2018github}, which we use extensively in our implementation. Section \ref{sec:jac} discusses how \texttt{JAX} gives an extra performance boost.

\begin{table}
	\centering
	\small
	\caption{Discretisation of PDAE from Section \ref{sec:p2d_pdae}}
	\label{tab:discretisations}
	\begin{tabular}{llr}
		\toprule
		Electrodes, $i \in \{p, n\}$
		\\ \midrule
		\addlinespace
		\DefEq{\textbf{Solid phase concentration}}{0 = c^{k-1}_{n,m} - c^k_{n,m} + D^s\left[ \frac{1}{r_n^2} \left( \frac{ r_{n+1/2}^2 \frac{c^k_{n+1,m} - c^k_{n,m}}{\Delta r} - r_{n-1/2}^2 \frac{c^k_{n,m} - c^k_{n-1,m}}{\Delta r}}{\Delta r} \right)\right] ,}\\
		\addlinespace
		\DisplayEq{\qquad \frac{c^k_{1,m} - c^k_{0,m}}{\Delta r} = 0, \qquad \frac{c^k_{N+1,m} - c^k_{N,m}}{\Delta r} + \frac{j^k_m}{D^s_{\eff}} =0.}
		%& \SetEqNum\label{eq:Volumen}
		\\
		\addlinespace
		\DefEq{\textbf{Electrolyte concentration}}{0 = u^{k-1}_m - u^k_m - \frac{\Delta t}{\epsilon_i}\left( \frac{{D_{\eff,m}}^r(u^k_{m+1} - u^k_{m})}{\Delta x^2} - \frac{{D_{\eff,m}}^l(u^k_{m} - u^k_{m-1})}{\Delta x^2} + a_i(1-t_+)j^k_m \right),}\\
		\addlinespace
		\DisplayEq{\frac{u^k_{p_1}-u^k_{p_0}}{\Delta x_p} = 0, \qquad \frac{u^k_{n_{M_n + 1}}-u^k_{n_{M_n}}}{\Delta x_n} = 0},\\
		\DisplayEq{\frac{u^k_{p_{M_p}}+u^k_{p_{M_p+1}}}{2} = \frac{u^k_{s_{0}}+u^k_{s_{1}}}{2}, \qquad \frac{u^k_{s_{M_s }}+u^k_{s_{M_{s+1}}}}{2} = \frac{u^k_{n_0}+u^k_{n_1}}{2}}.\\
		\addlinespace
		
		\DefEq{\textbf{Electrolyte potential}}{a_iFj^k_m + (\kappa^r_{\eff}\frac{\Phi^k_{m+1} - \Phi^k_m}{\Delta x^2} - \kappa^l_{\eff}\frac{\Phi^k_m - \Phi^k_{m-1}}{\Delta x^2}) - \gamma(\kappa^r_{\eff}T^k_{m+1}\frac{\ln{u^k_{m+1} - \ln{u^k_{m}}}}{\Delta x^2}) }\\
		\addlinespace
		
		\DisplayEq{\frac{\Phi_{{p_1}} -\Phi_{{p_0}} }{\Delta x} = 0, \qquad \frac{\Phi_{n_{N}} + \Phi_{n_{N+1}}}{2} = 0,}\\
		
		\DisplayEq{\frac{\Phi^k_{p_{M_p}}+\Phi^k_{p_{M_p+1}}}{2} = \frac{\Phi^k_{s_{0}}+\Phi^k_{s_{1}}}{2}, \qquad \frac{\Phi^k_{s_{M_s }}+\Phi^k_{s_{M_{s+1}}}}{2} = \frac{\Phi^k_{n_0}+\Phi^k_{n_1}}{2}}. \\
		\addlinespace
		
		\DefEq{\textbf{Solid phase potential}}{\sigma_{\eff,i} \frac{\Psi^k_{m+1} - 2\Psi^k_m + \Psi^k_{m-1}}{\Delta x_i^2}=0,}\\
		
		\DisplayEq{\sigma_{\eff,p} \frac{\Psi^k_{p_1} - \Psi^k_{p_0}}{\Delta x_p} = -I_{app}, \qquad \sigma_{\eff,p} \frac{\Psi^k_{p_{M_p +1}} - \Psi^k_{p_{M_p}}}{\Delta x_p} = 0, }\\
		\DisplayEq{\sigma_{\eff,n} \frac{\Psi^k_{n_1} - \Psi^k_{n_0}}{\Delta x_n} = 0, \qquad \sigma_{\eff,n} \frac{\Psi^k_{n_{M_n +1}} - \Psi^k_{n_{M_n}}}{\Delta x_n} = 0. }\\
		\addlinespace
		
	\end{tabular}	
\end{table}

\begin{table}
	\centering
	\small
	\caption{Discretisation of PDAE from Section \ref{sec:p2d_pdae}}
	\label{tab:discretisations_separator}
	\begin{tabular}{llr}
		\midrule
		Electrodes, $i\in \{p, n\}$
		\\ \midrule
		
		\DefEq{\textbf{Temperature}}{\rho_i C_{p,i} \frac{T^{k}_{m} - T^{k-1}_{m}}{\Delta t} -\left( \lambda_i\left[ \frac{T^k_{m+1} - 2T^k_m + T^k_{m-1}}{\Delta x ^2}\right]  + Q_{ohm} + Q_{rxn} + Q_{rev} \right) = 0,}\\ 
		\addlinespace
		\DisplayEq{-\lambda_a \frac{T^k_{a_{M_a+1}} - T^k_{a_{M_a}}}{\Delta x_a} = -\lambda_p \frac{T^k_{p_1} - T^k_{p_0}}{\Delta x_p}, \qquad -\lambda_n \frac{T^k_{n_{M_n+1}} - T^k_{n_{M_n}}}{\Delta x_n} = -\lambda_z \frac{T^k_{z_1} - T^k_{z_0}}{\Delta x_z}.}
		\\
		\DisplayEq{\frac{T^k_{p_{M_p}}+T^k_{p_{M_p+1}}}{2} = \frac{T^k_{s_{0}}+T^k_{s_{1}}}{2}, \qquad \frac{T^k_{s_{M_s }}+T^k_{s_{M_{s+1}}}}{2} = \frac{T^k_{n_0}+T^k_{n_1}}{2}}. \\
		\addlinespace
		
		\DefEq{\textbf{Ionic flux}}{j^k_m - 2_{k_\eff,i}\sqrt{u^k_m(c^{max} - c^{*k}_m)c^{*k}_m} \sinh{\frac{0.5F}{RT^k_m}\eta^k_m=0}}\\
		\addlinespace
		
		\DefEq{\textbf{Overpotential}}{\eta^k_m - \Psi^k_m + \Phi^k_m - U_i=0}
		\\
		\addlinespace
		\toprule
		Current collectors, $i \in \{a,z\}$
		\\ \midrule
		\DefEq{\textbf{Temperature}}{\rho_i C_{p,i} \frac{T^k_{m+1} - T^k_{m}}{\Delta t} -\left( \lambda_i\left[ \frac{T^k_{m+1} - 2T^k_m + T^k_{m-1}}{\Delta x ^2}\right]  +\frac{I^2_{app}}{\sigma_{\eff,i}} \right) = 0,}\\
		\DisplayEq{-\lambda_a\frac{T^k_{a_1} - T^k_{a_0}}{\Delta x_a} = h(T_{ref} - T^k_{a_{1/2}}), \qquad -\lambda_z\frac{T^k_{z_{M_z+1}} - T^k_{z_{M_z}}}{\Delta x_z} = h(T^k_{z_{M_z + 1/2}} - T_{ref}).} 
		\\ \midrule
		Separator, $i = s$
		\\ \midrule
		\addlinespace
		\addlinespace
		\DefEq{\textbf{Electrolyte concentration}}{0 = u^{k-1}_m - u^k_m - \frac{\Delta t}{\epsilon_i}\left( \frac{{D_{\eff,m}}^r(u^k_{m+1} - u^k_{m})}{\Delta x^2} - \frac{{D_{\eff,m}}^l(u^k_{m} - u^k_{m-1})}{\Delta x^2}\right),}\\
		\addlinespace
		\DisplayEq{-D^r_{\eff, p_{M_p}}\frac{u_{M_p+1} - u_{M_p}}{\Delta x_p} = -D^r_{\eff,s_0}\frac{u_{s_1} - u_{s_0}}{\Delta x_s}, \qquad -D^r_{\eff, s_{M_s}}\frac{u_{M_s+1} - u_{M_s}}{\Delta x} = -D^r_{\eff,n_0}\frac{u_{n_1} - u_{n_0}}{\Delta x_n}.}\\
		\addlinespace
		
		\DefEq{\textbf{Electrolyte potential}}{- (\kappa^r_{\eff}\frac{\Phi^k_{m+1} - \Phi^k_m}{\Delta x^2} - \kappa^l_{\eff}\frac{\Phi^k_m - \Phi^k_{m-1}}{\Delta x^2}) + \gamma(\kappa^r_{\eff}T^k_{m+1}\frac{\ln{u^k_{m+1} - \ln{u^k_{m}}}}{\Delta x^2}) = 0,}\\
		\addlinespace
		\DisplayEq{-\kappa^r_{\eff, p_{M_p}}\frac{\Phi_{M_p+1} - \Phi_{M_p}}{\Delta x_p} = -\kappa^r_{\eff,s_0}\frac{\Phi_{s_1} - \Phi_{s_0}}{\Delta x_s}, \qquad -\kappa^r_{\eff, s_{M_s}}\frac{\Phi_{M_s+1} - \Phi_{M_s}}{\Delta x_s} = -D^r_{\eff,n_0}\frac{u_{n_1} - u_{n_0}}{\Delta x_n}.}\\
		\addlinespace
		
		\DefEq{\textbf{Temperature}}{\rho_i C_{p,i} \frac{T^{k+1}_{m} - T^k_{m}}{\Delta t} -\left( \lambda_i\left[ \frac{T^k_{m+1} - 2T^k_m + T^k_{m-1}}{\Delta x ^2}\right]  + Q_{ohm} \right) = 0,} \\
		
		\DisplayEq{-\lambda_p \frac{T^k_{p_{M_p+1}} - T^k_{p_{M_p}}}{\Delta x_p} = -\lambda_s \frac{T^k_{s_1} - T^k_{s_0}}{\Delta x_s}, \qquad -\lambda_n \frac{T^k_{s_{1}} - T^k_{s_{0}}}{\Delta x_s} = -\lambda_n \frac{T^k_{n_{M_n + 1}} - T^k_{n_{M_n}}}{\Delta x_n}.}	
			\end{tabular}	
	
\end{table}

\section{Decoupled method}
As mentioned in Section \ref{sec:p2d_fd}, the major bottleneck of the computation comes from solving
the linear system 
\begin{equation}
\mathcal{J}(U^p) \delta = \bm{F}(U^p)
\end{equation}
at every Newton iteration at every time step. $\mathcal{J}$ is of size $\bigoh(NM) \times \bigoh(NM)$. In this section, we present a decoupled method to solve the linear system in two steps which results in a smaller Jacobian matrix of size $\bigoh(M) \times \bigoh(M)$. We do this by decoupling the equation for $\bm{c}$ from $\bm{F}(U)$. The technique results in an exact solve without any approximations. This way, we can reduce the computation cost of solving the linear system at every Newton iteration. From \eqref{eq:solid_conc_pde} we see that equation for lithium-ion concentration $c$ is linear except at the boundary of the sphere $r=R_p$. Writing the discretised \eqref{eq:solid_conc_pde} in a matrix form, we obtain
\begin{equation}\label{eq:ceq_matrix}
A\bm{c_m}^k = \begin{bmatrix}
0 \\ c_{1,m}^{k-1} \\ \vdots \\ c_{N,m}^{k-1} \\ 0
\end{bmatrix} - 
\frac{j_m^k}{D^s_{eff}(T^k_m)} \begin{bmatrix}
0 \\ 0 \\ \vdots \\ 0 \\ 1
\end{bmatrix},
\end{equation}
where  $\bm{c}^k_m$ is the particle equation at point $x_m$; therefore, for each $m = 1, 2, \dots, M$, we would have to solve \eqref{eq:ceq_matrix}. Without the decoupling, which will be described shortly, \eqref{eq:ceq_matrix} is solved as part of the larger system, as it is coupled to every other variable through $j$. $A$ is the tridiagonal coefficient matrix for and $\bm{c}^k_m$. Let $ a_n= \frac{-D^s\Delta t}{r^2_{n}\Delta r^2}r^2_{n-1/2}$, $b_n= 1+ \frac{D^s\Delta t}{r^2_{n}\Delta r^2}$, $c_n=\frac{-D^s\Delta t}{r^2_{n}\Delta r^2}r^2_{n+1/2}$. Then, 
\begin{equation} \label{eq:A}
A = \begin{pmatrix}
-1 & 1 & 0 & 0 &  0 & \dots & 0\\
0 & a_1  & b_1  & c_1 & 0 & \dots & 0\\
0 & 0  & a_2  & b_2 & c_2 & \dots & 0\\
\vdots &\vdots& \ddots & \ddots & \ddots & \ldots & \vdots\\
0 & 0 & \ldots & \ldots & \ldots & \frac{-1}{\Delta r} & \frac{1}{\Delta r}
\end{pmatrix}.
\end{equation}
We now show how the system can be decoupled. Rearranging \eqref{eq:ceq_matrix}, we can split $\bm{c}^k_m$ into two parts
\begin{align*}
\bm{c}^k_{I_m} = A^{-1} \begin{bmatrix}
0 & c^{k-1}_{0,m} & \ldots & c_{N,m}^{k-1} & 0
\end{bmatrix}^T, \\
\bm{c}^k_{II_m} = \frac{j_m^k}{D_s(T^k_m)} A^{-1} \begin{bmatrix}
0 & 0 & \ldots & 0 & 1
\end{bmatrix}^T.
\end{align*}
The constant vector $A^{-1} \begin{bmatrix}
0 & 0 & \ldots & 0 & 1
\end{bmatrix}^T$ in $\bm{c}^k_{II_m}$, which we denote $\gamma$, is a vector that can be precomputed once. By formulating as such, we have partially decoupled the particle equations from the whole system. This is because we can compute $\bm{c}^k_{I_m}$ by solving
\begin{equation} \label{eq:tridiag_system}
A\bm{c}^k_{I_m} = \begin{bmatrix}
0 \\ c_{1,m}^{k-1} \\ \vdots \\ c_{N,m}^{k-1} \\ 0
\end{bmatrix}
\end{equation}
cheaply via tridiagonal LU factorisation. Thus, we avoid the full $\bigoh(NM) \times \bigoh(NM)$ Jacobian of $F(U)$. The vector $\bm{c}^k_{II_m}$ can be obtained by scalar-vector multiplication once we have a solution for $\bm{j}$ and $\bm{T}$. Equations that are functions of $\bm{c}$, \eqref{eq:p2d_j}, \eqref{eq:p2d_T} and \eqref{eq:p2d_eta}, only require the interfacial concentration $c^*(x,t)$. 
In our original system, $c^*$ is approximated as
\begin{align*}
c^*(x_m, t)  \approx c^{*^k}_m &= \frac{c^k_{N,m} + c^k_{N+1,m}}{2}.
\end{align*}
This can be rewritten as
\begin{align*}
c^{*^k}_m &= \underbrace{\frac{c^k_{1_{N,m}} + c^k_{1_{N+1,m}}}{2}}_{ \textstyle =\alpha^k_m} - \frac{j^k_m}{D^s_{\eff}(T^k_m)}\underbrace{\frac{\gamma_{N+1} + \gamma_N}{2}}_{\textstyle =\beta}.
\end{align*}
Treating $\alpha^k_m$ and $\beta$ as constants, $c^*(x,t)$ is a function of $j$ and $T$, $c^{*k}_m = \alpha^k_m - \frac{j^k_m}{D^s(T^k_m)}\beta$. In this way, none of the channel equations depend explicitly on variable $\bm{c}$. Therefore, $\bm{c}$ does not need to be computed in the Newton iterations. We can proceed with the Newton iteration with $U = (\bm{u}, \bm{j}, \bm{\eta}, \bm{T} \bm{\phi}, \bm{\psi})$ where we solve a linear system of a $\bigoh(M) \times \bigoh(M)$ matrix. Table \ref{tab:num_grid} compares the size of the Jacobian for each method. We stress here that the decoupled method solves the same system as the na\"{i}ve method. We obtain the same solution to machine precision. If there is a nonlinear dependence in the particle  diffusion, the decoupling will still apply, although the tridiagonal solves \eqref{eq:tridiag_system} for each particle will need to be updated at each Newton step.

In Section \ref{sec:reordered}, we further improve the efficiency by reordering the entries of Jacobian and computing only the nonzero entries of the reduced Jacobian. 

\begin{table}[htp]
	\centering
	\begin{tabular}{c|c|c|c|c}
		$N_p = N_n$ & $M_p =M_n$ & $M_s=M_a=M_z$ & $N_{\text{jac}}$: naive & $N_{\text{jac}}$: decoupled \\
		\hline
		10 & 10 & 5 & 411 & 171\\
		20 & 20 & 5 & 1171 & 291 \\
		30 & 30 & 5 & 2331 & 411 \\
		40 & 40 & 5 & 3891 & 531 \\
		50 & 50 & 5 & 5851 & 651
	\end{tabular}
	\caption{The table shows the size of the Jacobian for each method at different resolutions. The same number of grid points in $r$ and $x$ for positive and negative electrode was used. The number of grid points was set fixed to $5$ in the separator, positive and negative current collector. The size of the Jacobian grows by $\bigoh(NM)$ for the naive method, whereas only $\bigoh(M)$ for decoupled.}
	\label{tab:num_grid} 
\end{table}

\subsection{Reordering the Jacobian}\label{sec:reordered}
The vector $U$ from our decoupled system can be reordered such that the Jacobian is a banded diagonal matrix. By reordering, we can use an appropriate banded linear solver to enhance the performance. We rearrange $U$ in an alternating fashion,
%\begin{split}
%	U &= 
%	[\begin{matrix} c_p & c_n & u_p & u_s & u_n& j_p& j_n& \eta_p & \eta_n \end{matrix} \\
%	&\qquad\qquad \begin{matrix}  \Phi_p & \Psi & \Phi_n & \Psi_p & \Psi_n & T_a & T_p & T_s & T_n & T_z \end{matrix}]^T.
%\end{split}
\begin{equation}\label{eq:reorder_full}
\begin{split}
U &=
[\begin{matrix}
\bm{u}_{p}  & \bm{u}_s & \bm{u}_n & \bm{j}_p & \bm{j}_n & \bm{\eta}_p & \bm{\eta}_n \end{matrix} \\
&\qquad\qquad \begin{matrix} \bm{\Psi}_p &\bm{\Psi}_n& \bm{\Phi}_p &\bm{\Phi}_s &\bm{\Phi}_n & \bm{T}_a & \bm{T}_p & \bm{T}_s& \bm{T}_n&\bm{T}_z
\end{matrix}]^T
\end{split}
\end{equation}
$$\downarrow$$
\begin{equation}
\begin{split}
U_r&=[ 
\begin{matrix}
\bm{T}_a &
{u}_{p_0} & 
{\Psi}_{p_0} &
{\Phi}_{p_0} & 
{T}_{p_0} \end{matrix}\\
&\qquad \qquad \begin{matrix} 
u_{p_1} & j_{p_0} & \eta_{p_0} & \Phi_{s_{p_1}} & \Phi_{e_{p_1}} & T_{p_1} & u_{p_2} \cdots & T_{p_{Mp}}
\end{matrix} \\
&\qquad \qquad \qquad \begin{matrix}
{u}_{p_{Mp+1}}
{\Psi}_{p_{Mp+1}}&
{\Phi}_{p_{Mp+1}} &
{T}_{p_{Mp+1}} \end{matrix}\\
&\qquad  \qquad \qquad \qquad \begin{matrix}
{u}_{s_0}&
{\Phi}_{s_0}&
{T}_{s_0} & u_{s_1} & \Phi_{e_{s_1}} & T_{s_1} & \cdots & 
{u}_{s_{Ms+1}}&
{\Phi}_{s_{Ms+1}}&
{T}_{s_{Ms+1}}\end{matrix}\\
&\qquad  \qquad \qquad  \qquad \qquad \begin{matrix}
\cdots & u_{n_1} & j_{n_0} & \eta_{n_0} & \Phi_{s_{n_1}} & \Phi_{e_{n_1}} & T_{n_1}&\cdots  & \bm{T}_z
\end{matrix}]^T
\end{split}
\end{equation}

\begin{figure}
	\begin{minipage}{2.5in}
		\includegraphics[scale=0.3]{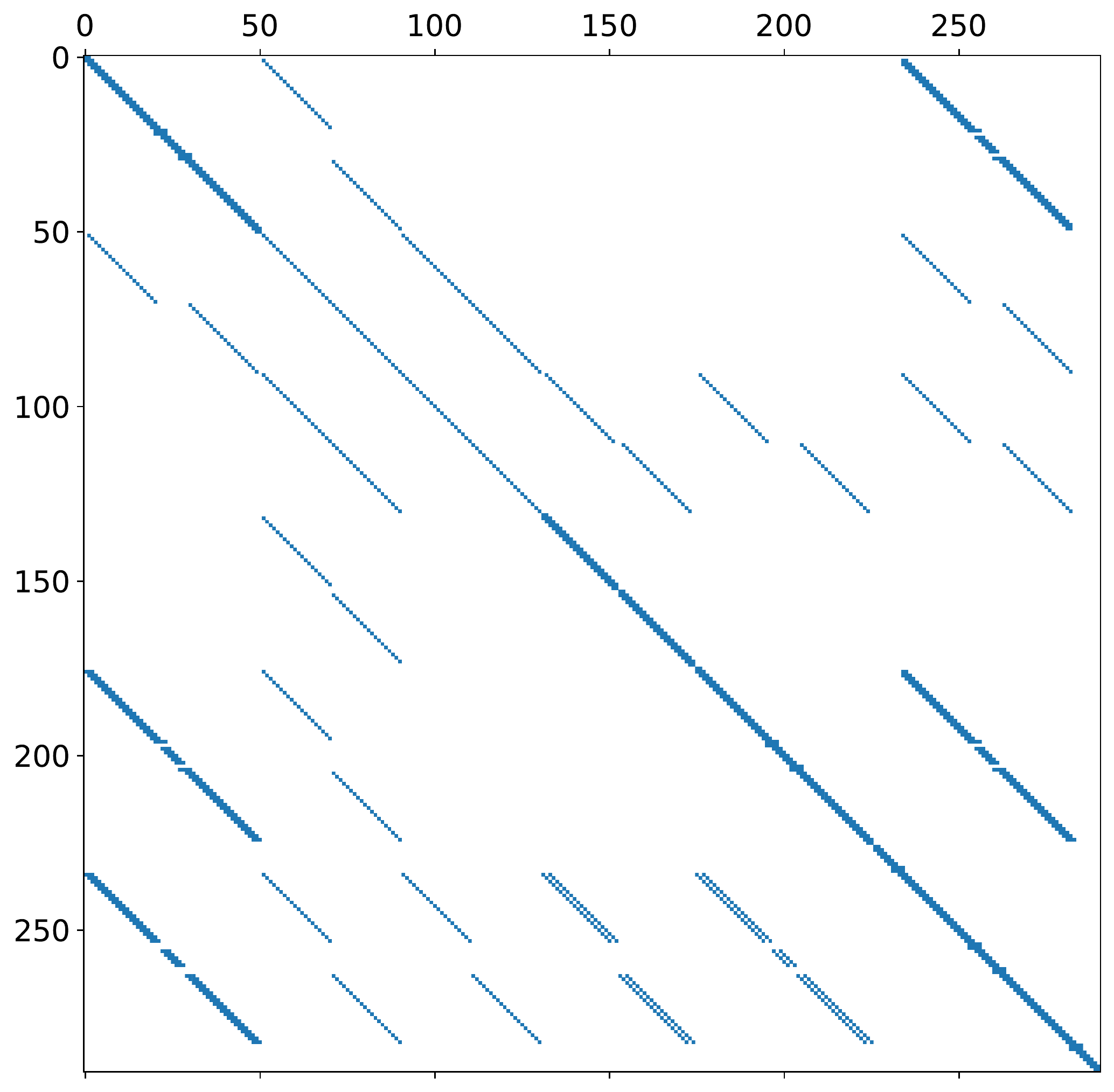}
	\end{minipage}
	\begin{minipage}{2.5in}
		\includegraphics[scale=0.3]{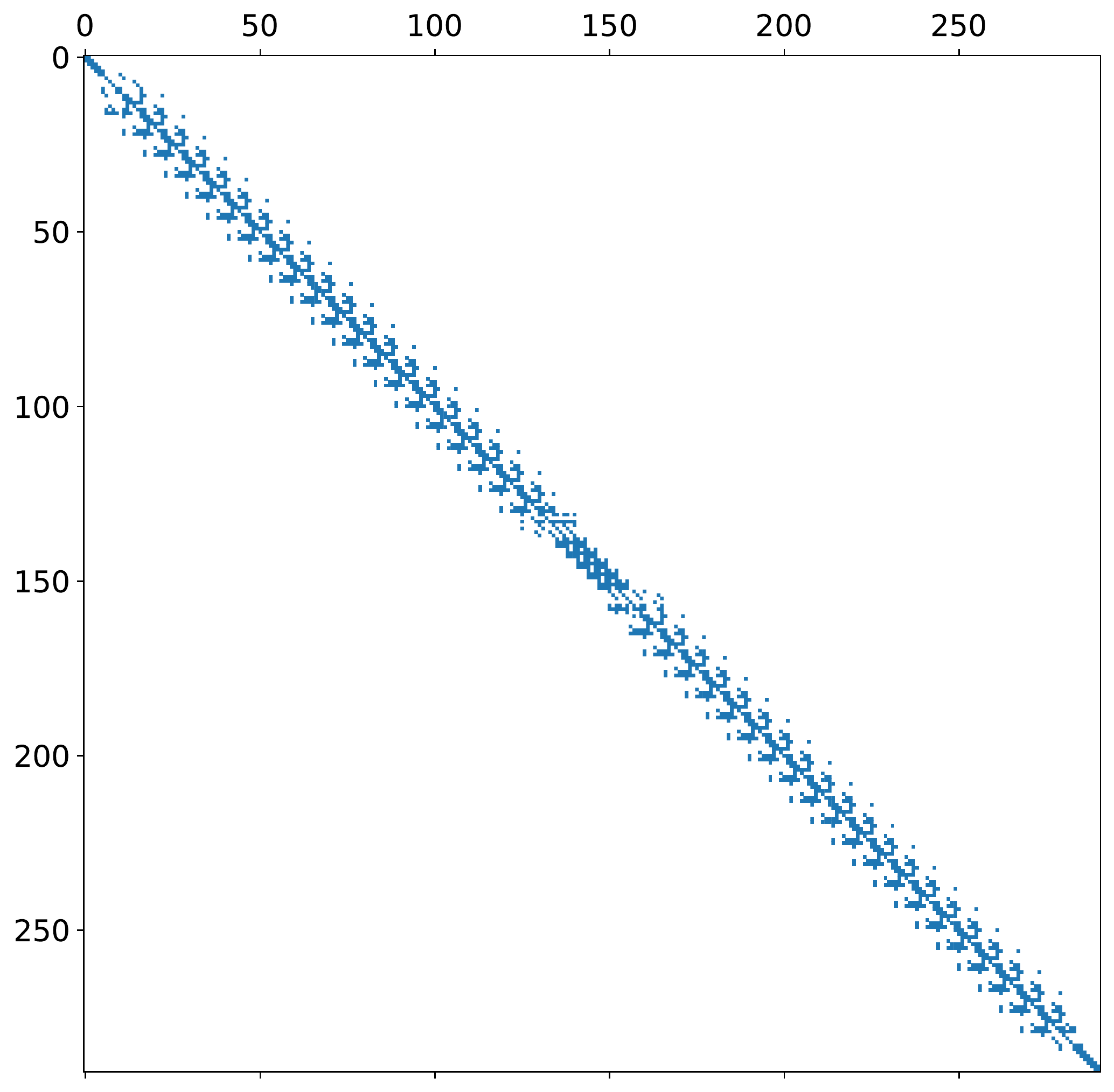}
	\end{minipage}
	\caption{The sparsity pattern of Jacobian from the unordered system on the left. On the right is the sparsity pattern after the Jacobian is reordered.}
	\label{fig:sparsity}
\end{figure}

The variables are ordered such that $u_i, j_i, \eta_i, \Psi_i, \Phi_i$ and $T_i$ at each $i-$th grid point are next to each other, rather than the original set up described in \eqref{eq:bigU}. Figure \ref{fig:sparsity} illustrates how the sparsity pattern of the Jacobian changes.

\subsubsection{Banded matrix}\label{sec:banded_matrix}
The reordered Jacobian, which we denote by $\mathcal{J}_r$, can be stored in a diagonal ordered form. There are $23$ nonzero lower and $23$ upper diagonals in $\mathcal{J}_r$. Then, $\mathcal{J}_r$ can be represented as a matrix of size $(23 + 23 + 1) \times N_{tot}$ where $N_{tot}$ is the size of $U$. We denote this banded matrix by $\mathcal{J}_b$. The entries are stored in the following way,
\begin{gather*}
\mathcal{J}_b[u + i - j, j] = \mathcal{J}_r[i,j], \quad  i = 0, 1, \cdots , (l + u), \quad j = 0, 1, \cdots, N_{tot}.
\end{gather*} 
The number of rows in $\mathcal{J}_b$ is fixed for a given model, as the number of rows correspond to the number of partial derivatives in the system. $\mathcal{J}_b$ is passed to the banded linear solver, which uses $LU$ decomposition to solve the linear system.

\subsection{Constructing the Jacobian}\label{sec:jac}
We describe some of the performance gains that can be achieved by using \texttt{JAX} in Python. \texttt{JAX} is especially useful for high-performance computations because it uses a domain-specific compiler \texttt{XLA} (Accelerated Linear Algebra) that can optimize computations at a lower level with improved memory usage and speed. Compiling code just-in-time (jit) via \texttt{XLA} in Python accelerates the performance because the source code is directly compiled to machine code as the program runs. This is often faster than interpreting, which is typical for languages like Python, making the performance comparable to that of compiled languages like C. The readers can refer to \cite{jax2018github} and \cite{xla} for more information. 

In our problem, a fast compiler that optimizes differentiation computation is very useful because we need to compute the system Jacobian at $U$ at every Newton step. In addition to differentiation, the compiler can speed up any array manipulations. Any functions that are compiled using \texttt{XLA} incur initial overhead when first called, and is cached to be used later. This is useful for functions that are called many times, as is the case in our problem with multiple Newton updates and time stepping. For example, the first time we build the Jacobian, there is an overhead of roughly 15 seconds (independent of the size of the vector). The proceeding computations are fast.

However, there are limitations of using \texttt{JAX}'s built-in forward-mode Jacobian function. While \texttt{JAX} can automatically output the Jacobian when $U$ is fed into its built-in function, the output is dense. Currently, the sparse output of Jacobians is not supported. This creates an extra overhead of converting the matrix into a sparse form to use sparse linear solvers. This drawback also causes a computational bottleneck. The evaluation of Jacobian with \texttt{JAX}'s built-in function is quite costly for large vectors, even when compiled with \texttt{XLA}. This is likely coming from the redundant computation effort computing partial derivatives that evaluate to zeros. By the method discussed in Sections \ref{sec:reordered}, we bypass this problem by directly constructing the banded reordered Jacobian, $\mathcal{J}_b$, without using \texttt{JAX}'s built-in Jacobian function.

For each discretised PDE equation, we take the partial derivatives with respect to their arguments via backward mode AD on a scalar level ($\mathbb{R}^{\text{Ntot}} \rightarrow \mathbb{R}$). This avoids computing redundant zeros. Many of the discretised equations are repeated over the grid points, but they can be vectorized. We predetermine the sparsity of $\mathcal{J}_b$ prior to the simulation, which is unique to the reordering scheme and the model. At each Newton update, we populate the matrix with partial derivatives with the predetermined sparsity. The array update process is sped up by compilation via \texttt{XLA}. This method provides an efficient way constructing the Jacobian that takes advantage of the tools provided by \texttt{JAX}.

\section{Experimental setup}
All simulations were done on Mac CPU, with 3.5 GHz Dual-Core Intel Core i7 processor and 8GB RAM, in Python. Our code is freely available on Github at \url{https://github.com/hanrach/p2d_fast_solver}.
We ran the simulations until the full battery discharge at $1$C, which is observed at $t=3620$ seconds. At the current collectors and the separator, coarse grids ($M_a, M_s, M_z = 5$) were used for speed, because of the simpler governing equations in those sections. We varied the grid sizes in the electrodes, with same number of grid points at each positive and negative electrode in $x$ and $r$ ($M_p=N_p=M_n=N_n$). The time step in the backward Euler method was fixed to $\Delta t = 10$.

%For the \textit{fast solver}, which will be described in \ref{sec:comparison_naivefast}, we used same grid resolutions across the battery sections as the performance was similar.

\subsection{Voltage curves under different C rates}
\begin{figure}[htp]
	\centering
	\includegraphics[scale=0.7]{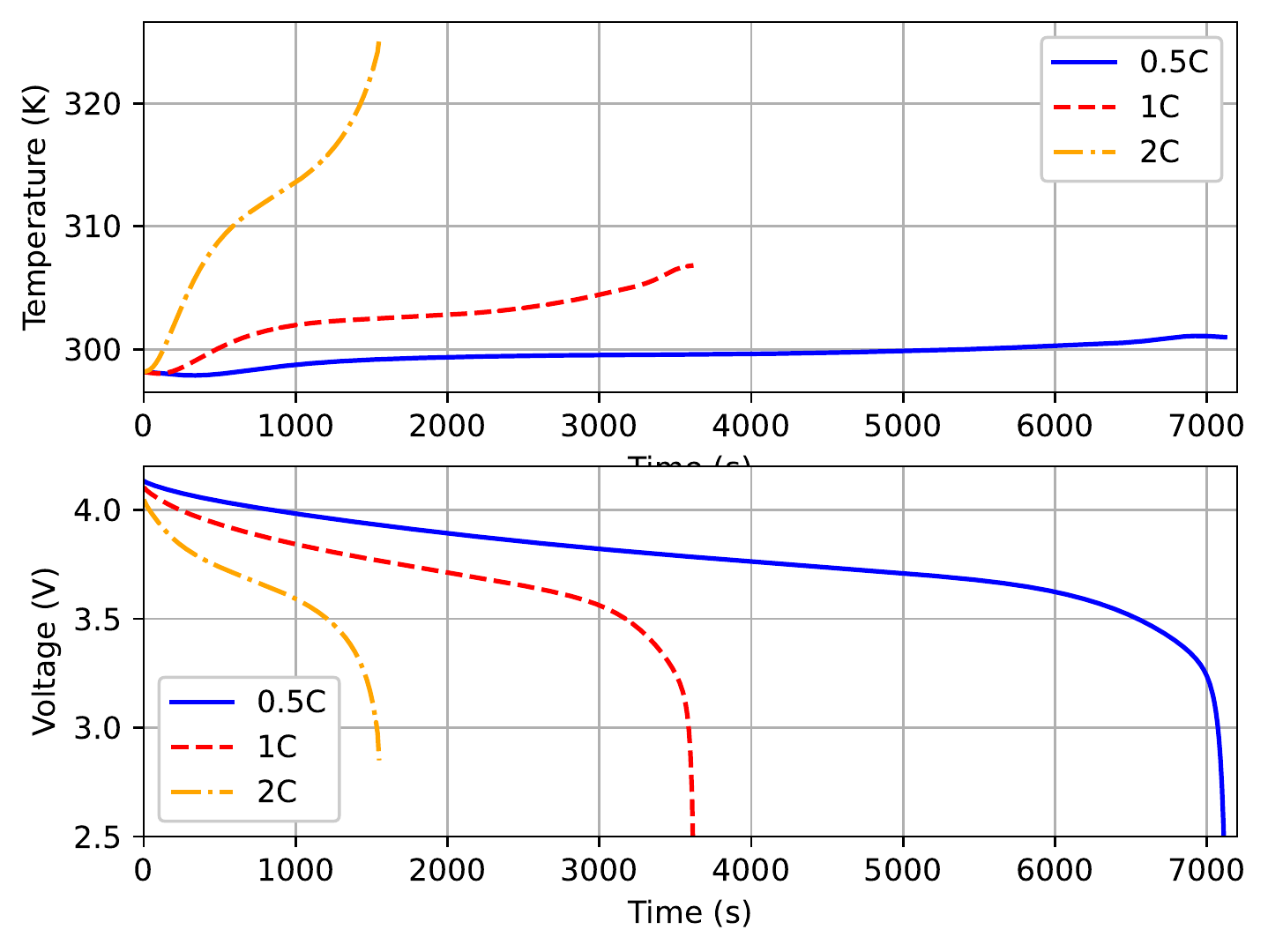}
	\caption{Temperature and voltage changes during the discharge cycle under 0.5C, 1C and 2C.}
	\label{fig:diff_c_rates}
\end{figure}

We experimentally validate our solver by comparing the voltage curves against Figure 9 in \cite{torchio2016}. We run the simulation until the full discharge at different C rates with the heat exchange coefficient $h$ fixed to $1$W/(m$^2$K). We plot how the average temperature and the battery voltage changes over time in Figure \ref{fig:diff_c_rates}. The battery voltage is defined as the difference in solid phase potential at the two ends of the electrodes,  $V = \Psi_{p_0} - \Psi_{n_{M_n}}$. As expected, at higher C rates the temperature rises rapidly and the voltage drops quickly to the cut-off voltage of $2.5$V. These figures are qualitatively similar to Figure 9 in \cite{torchio2016}. The script \verb|examples/different_C_rates.py| was used to generate Figure \ref{fig:diff_c_rates}.

\subsection{Comparsion between the na\"ive method and the fast solver}\label{sec:comparison_naivefast}
We compare the performance between the \textit{na\"ive method}, which does not include the decoupling and uses the built-in Jacobian function in \texttt{JAX}, and the \textit{fast method} where we decouple $c$ as well as construct the Jacobian in a diagonal ordered form directly. When solving the linear system $J\delta^p = \mbf{U}^p$ for the {na\"ive} method, we convert the dense Jacobian into a sparse matrix and apply a generic sparse linear solver. Table \ref{tab:methods_summary} summarizes the methods. 

\begin{table}[htp]
	\centering
	\begin{tabular}{c|c|c|c}
		& Decoupled & Jacobian Function & Linear Solver\\
		\hline 
		Na\"ive &  No & \texttt{JAX} & Generic sparse \\
		\hline
		Fast &  Yes&   Custom & Tridiagonal and banded \\
		
	\end{tabular}
	\caption{The fast method improves upon the na\"ive method by decoupling the solid particle equations from the electrolyte equations, and employing the custom Jacobian function and linear solvers that exploit the structure of the matrices.}
	\label{tab:methods_summary} 
\end{table}  

\begin{figure}
	\centering
	\includegraphics[scale=0.6]{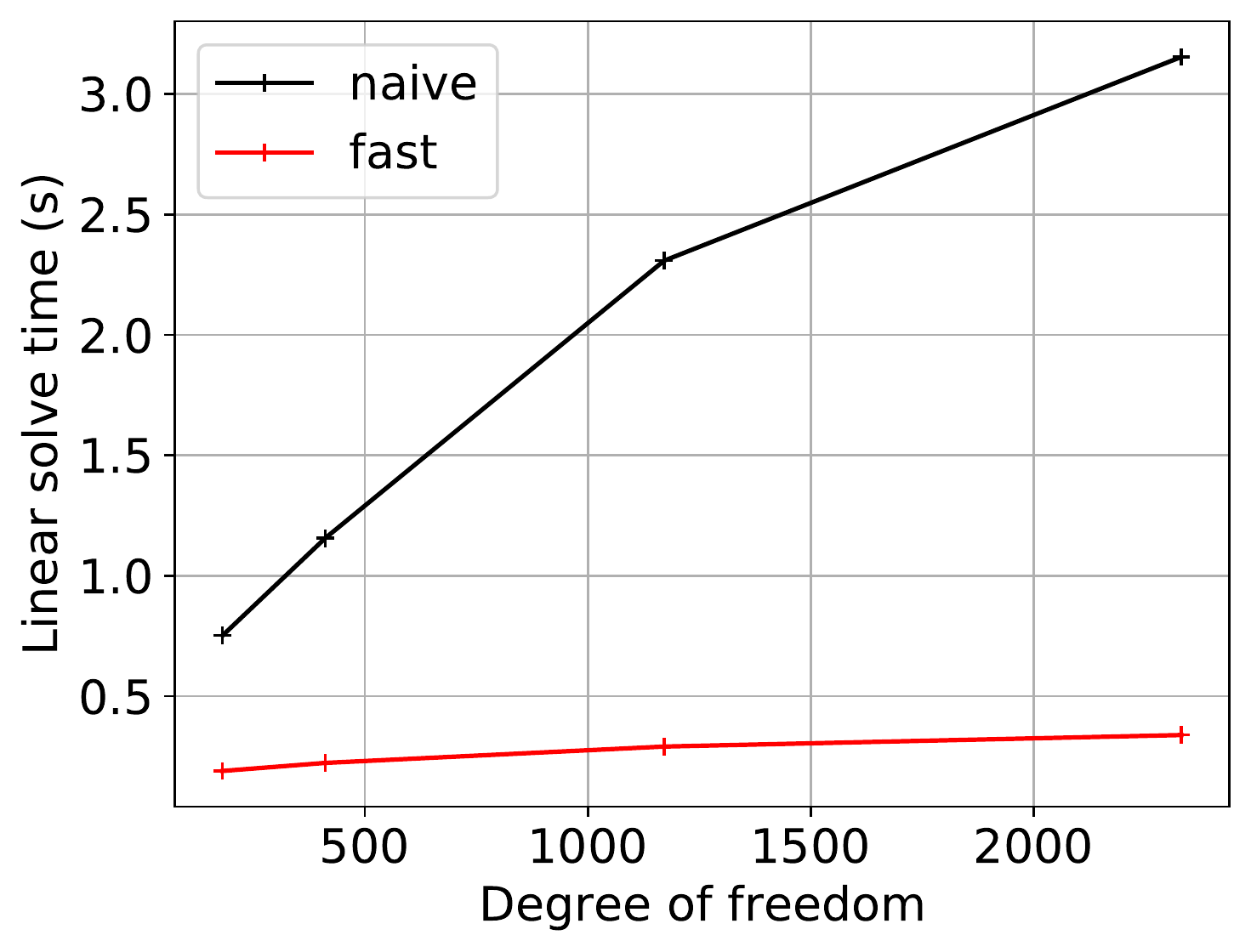}
	\caption{The total time of linear solves after the solver is run up to $t=3520$. The $x$-axis corresponds to the size of the solution (particle and channel). $M=N$ tested are $5, 10, 20$ and $30$. The time to assemble the coefficients of the linear systems \eqref{eq:ceq_matrix} are not included, since they can be precomputed. }
	\label{fig:full_linsolve}
\end{figure}

Figure \ref{fig:full_linsolve} compares the cumulative time spent on solving the linear systems within the Newton iterations. The overhead of conversion for the na\"ive method is not costly compared to the actual time to solve the linear systems. This factor is not included in the measurements in Figure \ref{fig:full_linsolve}. At a moderately fine grid $M=30$, the na\"ive method spends a total of about $3.5$ seconds. The fast solver is roughly $12$ times faster than the na\"ive method, spending about $0.3$ seconds.

\begin{figure}
	\centering
	\includegraphics[scale=0.7]{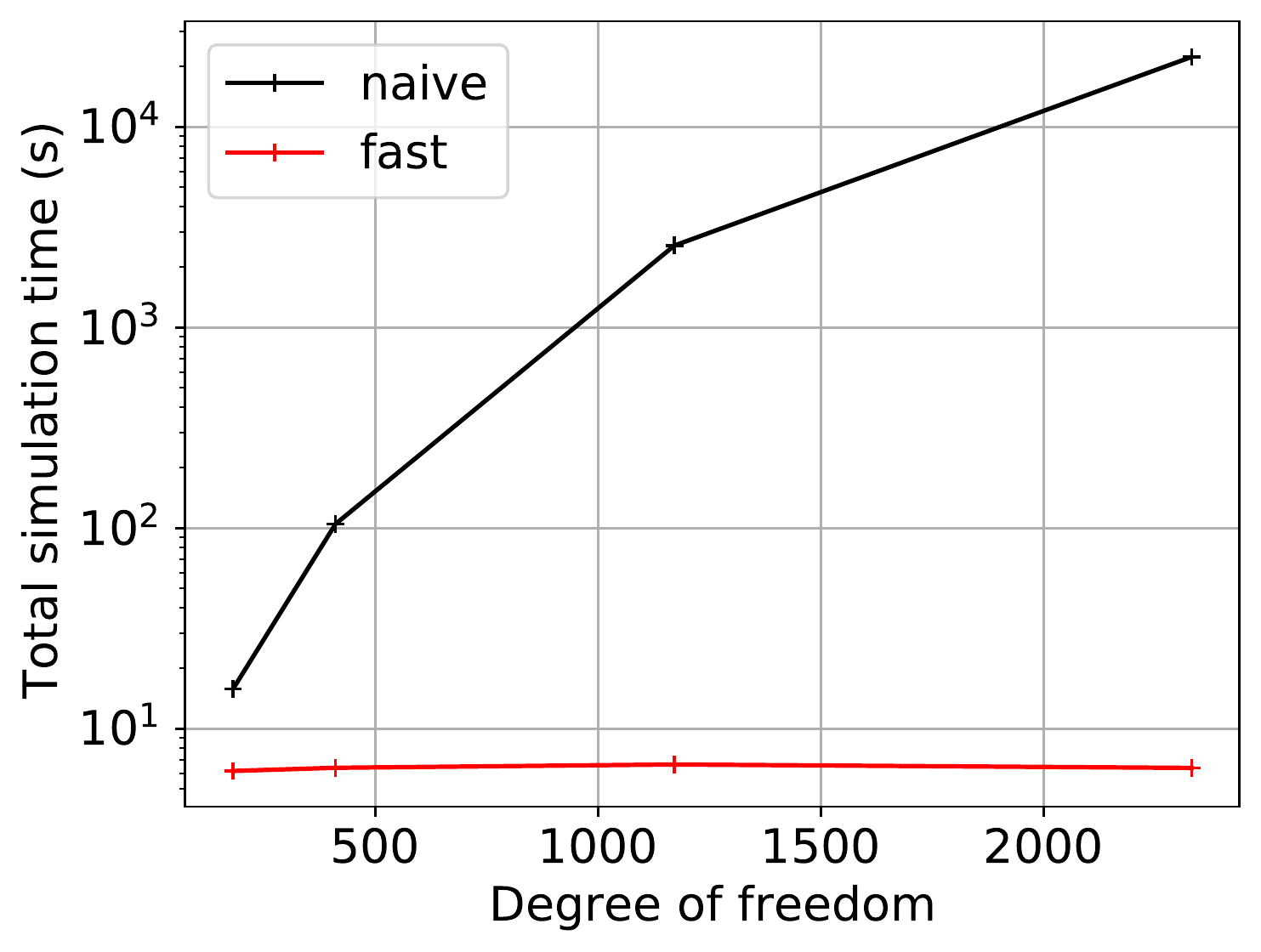}
	\caption{The total simulation time until full discharge at $1$C. $M=N$ tested are $5, 10, 20$ and $30$. }
	\label{fig:tot}
\end{figure}

Figure \ref{fig:tot} compares the total time of simulation between the naive solver and the fast solver for varied $M=N$. We especially observe that the time spent on evaluating the Jacobian is drastically reduced. At $M=30$, the fast solver outperforms the naive method by $\approx 5000$ times. 

The code used to generate the performance measurements can be found in the script \verb|performance/full_simulation_comparison.py|. The data in Table \ref{tab:compare} was also generated by 
the same code. The script \verb|performance/reordered_performance.py| was used to obtain data for Table \ref{tab:reorder_data} and Table \ref{tab:solvers_compare}.

\subsection{Comparison between the fast solver and other solvers}
We also compare the solver against \texttt{LIONSIMBA} \cite{torchio2016}, \texttt{DUALFOIL} \cite{Doyle_1993} and \texttt{COMSOL} \cite{COMSOL}. We take the P2D model simulation data of
\texttt{LIONSIMBA}, \texttt{DUALFOIL} and \texttt{COMSOL} from \cite{torchio2016}, which simulates a discharge cycle of 1C under isothermal conditions, done on a Windows 7 PC with 8GB of RAM.
We also include the runtimes of \texttt{LIONSIMBA} obtained on our machine (3.5 GHz Dual-Core Intel Core i7 with 8GB of RAM).
For these simulations, the grid size was kept the same across all battery sections.
Table~\ref{tab:solvers_compare} shows that the simulation times from our fast solver stay roughly constant for the resolutions tested, whereas this is not true for other solvers.

\begin{table}[htp]
	\begin{tabular}{@{}llllll@{}}
		$M$& 10     & 20     & 30     & 40     & 50     \\
		\toprule
		Fast solver$^\dagger$    & 4.09 s& 4.24 s &3.98 s& 4.15 s& 4.49 s \\
		\texttt{LIONSIMBA}$^\dagger$ & 7.46 s & 9.85 s & 15.48 s & 26.80 s & 54.60 s\\
		\toprule
		\texttt{LIONSIMBA}*     & 28 s  & 57 s  & 105 s  & 134 s  & 223 s  \\
		\texttt{DUALFOIL}*    & 28 s  & 69 s  & 97 s  & 137 s  & 185 s  \\ 
		COMSOL*    & 96 s  & 114 s  & 143 s  & 189 s  & 244 s \\ 
	\end{tabular}
	\caption{Comparison of simulation time between solvers. $^\dagger$The data was obtained from 3.5 GHz Dual-Core Intel Core i7 with 8GB of RAM. *The data was taken from \cite{torchio2016} collected on a different machine. }
	\label{tab:solvers_compare}
\end{table}
\section{Results}

The fast method improves the efficiency of the naive solver in two ways. Firstly, by decoupling the system, we reduce the time to solve the linear systems, as shown in Figure \ref{fig:full_linsolve}. The figure compares performance between the general sparse linear solver and the banded solver (\texttt{scipy.linalg.solve\_banded}) along with tridiagonal solves with $LU$ factorisation. 
By reformulating the problem and taking advantage of the linear structure in \eqref{eq:ceq_matrix}, we improve the linear algebraic efficiency.

Secondly, we work around the major bottleneck of the computation by implementing our own Jacobian building function. Under the AD framework with \texttt{JAX}, a large portion of the simulation time is spent on evaluating the Jacobian and $\bm{F}(U)$. This dominates the total simulation time shown in Figure \ref{fig:tot}. \texttt{JAX}'s Jacobian function takes up to six hours to evaluate the Jacobians with the full channel and particle system. Decoupling alone without using our custom Jacobian building function (the \textit{decoupled method}) significantly cuts the time to evaluate the Jacobian. Evaluating the Jacobian of size $\bigoh(M) \times \bigoh(M)$ is much cheaper with \texttt{JAX}'s Jacobian function, as opposed to $\bigoh(NM) \times \bigoh(NM)$. For all grid points tested, the decoupled method is faster than the naive method, drastically outperforming as $M$ increases. The decoupled formulation achieves performance that is $\approx870$ times faster than the na\"ive method at $M=30$. 

The fast method further improves the efficiency. \texttt{JAX} is only used to compute the partial derivatives via backward mode differentiation. Compared to the decoupled method, the custom function reduces the time to evaluate the Jacobian up to $\approx7$ times at $M=30$. This is evidently due to the fact that the banded matrix in diagonal ordered form (of size $47 \times N_{tot}$) has a fixed number of rows and only the number of columns increases with $M_a, M_p, M_s,M_n$ and $M_z$. Table \ref{tab:compare} compares the time to auto differentiate and evaluate $\mathcal{J}(U)$ and $\bm{F}(U)$ and the total simulation time of the three methods discussed.

\begin{table}[htp]
	\begin{tabular}{@{}lll@{}}
		& Evaluation (s) & Total (s)   \\
		\toprule
		Naive    & \num{2.22e4} s  & \num{2.31e4} s    \\
		Decoupled only (\textit{decoupled method})  & 15.72 s  & 21.63 s    \\
		Decoupled \& Banded (\textit{fast method})   & 0.3943 s  & 3.898 s  \\ 
	\end{tabular}
	\caption{Evaluation time is the time spent on evaluating $\mathcal{J}(U)$ and $\bm{F}(U)$. The simulations were compared at $M=30$.}
	\label{tab:compare}
\end{table}

Table \ref{tab:reorder_data} summarises the performance of the fast solver. There is an initial overhead ($\approx 15$ seconds) that comes with using \texttt{XLA} compiler, which can be precomputed before the actual simulation.  Therefore, we do not factor it into the total time. We observe that the total time spent on the simulation stays roughly constant between three to five seconds. The grids at all battery domains were set equal, with $N_p = N_n = 10, 20, 30, 40, 50$ and $M_a = M_p = M_s = M_n = M_z =10, 20, 30, 40, 50$. For these grid resolutions, we observe that the evaluation and solve times are not the dominating factors but  the miscellaneous calculation overhead. This explains why the simulation time seem constant with respect to $N$ and $M$. Choosing $M=N \gtrapprox 100$, a clearer upward trend emerges where the solve time dominates the computation (results not included). For practical applications, such fine resolution may not be needed and $M=N\leq50$ would suffice.
\begin{table}[htp]
	\begin{tabular}{@{}llllll@{}}
		$M$& 10     & 20     & 30     & 40     & 50     \\
		\toprule
		Solve    & 0.213 & 0.289 & 0.346& 0.4174& 0.532  \\
		Evaluation     & 0.376 &0.412& 0.386& 0.426& 0.483 \\
		Total    & 4.092& 4.243 &3.987& 4.1536& 4.494 \\ 
		%		Eval     & 2.453  & 2.483  & 2.742  & 2.830  & 2.625  \\
		%Initial  & 13.909 & 14.757 & 16.479 & 16.058 & 14.412 \\ \bottomrule
		%		Overhead & 0.200  & 0.205  & 0.206  & 0.217  & 0.202 \\ \bottomrule
	\end{tabular}
	\caption{Averaged times of the fast solver over 20 runs. Solve time indicates the total time spent on linear solves. Evaluation time is the time spent on evaluating $\mathcal{J}(U)$ and $\bm{F}(U)$.}
	\label{tab:reorder_data}
\end{table}

\section{Conclusion}

We have implemented a fast and robust implementation of the full P2D model without reducing the model. We give a thorough description of the discretisation of the PDAE involving cell-centred finite differences and backward Euler time stepping. The resulting system is a set of nonlinear equations, which can be solved by Newton's method. 

In order to robustly compute the Jacobian at every Newton's iteration, automatic differentiation is used via the Python package \texttt{JAX}. Because the full P2D model has a linear structure in the particle equations \eqref{eq:solid_conc_pde}, we are able to decouple the particle and channel equations at each time step. After decoupling, we solve one large tridiagonal linear system per time step and smaller sparse systems (channel equations) within the Newton iterations. The full system is thus solved more efficiently.

The decoupled method enables us to run the simulation 1C discharge cycle in a reasonable time under the AD framework, whereas such framework is infeasible when we use the naive method. This was made possible by rearranging the variables in the channel equations to create a matrix that is banded, which can be solved efficiently. The computation was made more efficient by directly building the banded Jacobian using the \texttt{XLA} compiler. 

An additional way we can increase the efficiency is implementing a adaptive time stepping scheme, based on error on $\bm{c}$ only. We have implemented this for the scaled P2D model based on \cite{scaledp2d}, which is a work in progress by Hennessey et al. The adaptive time stepping takes larger time steps in the beginning and gradually adjusts to smaller steps as the problem gets harder near the end of the cycle. For the scaled P2D model, it improved the efficiency by roughly a factor of two compared to the fixed time stepping scheme. This is future work to be explored.

%A weakness of the solver is that the sparsity of the banded structure is fixed for a given reordering scheme and the model, which reduces the flexibility of the solver. For example, if the dependencies on variables change in  equations, we would have to reflect the change in the Jacobian sparsity pattern in the code. Improving the process of building the banded Jacobian remains as future work to be explored. 

We have demonstrated that by exploiting the structure of the problem, and utilizing state-of-art software, the complex P2D model can be solved accurately in under a few seconds. The fast solver can be potentially used as part of larger computations, such as optimizing multiple-battery pack behaviours. The solver would be especially useful where the P2D model has to be solved repeatedly over many cycles for estimation, optimization or control purposes. 
\appendix

\section{Symbols and Additional equations }\label{sec:appen_eqn}
\printunsrtglossary[title={List of Symbols},type=symbols,style=long]

\begin{center}
	\footnotesize
	\begin{longtable}{l}
		Open Circuit Potential\\
		\hline\\
		$\bm{U}_p = \bm{U}_{\text{p,ref}} + (T(x,t) - T_{\text{ref}}) \frac{\del \bm{U}_p}{\del\bm{T}}\rvert_{T_\text{ref}}$ \\
		\\
		$\bm{U}_n = \bm{U}_{\text{n,ref}} + (T(x,t) - T_{\text{ref}}) \frac{\del \bm{U}_n}{\del\bm{T}}\rvert_{T_\text{ref}}$\\
		\\
		Entropy Change \\
		\hline \\
		$\frac{\del \bm{U}_p}{\del\bm{T}}\rvert_{T_\text{ref}} = -0.001\left( \frac{0.199521039 - 0.928373822\theta_p + 1.364550689000003\theta_p^2  - 0.6115448939999998\theta_p^3}{1 - 5.661479886999997\theta_p + 11.47636191\theta_p^2 - 9.82431213599998\theta_p^3 + \
			3.046755063\theta_p^4} \right)$ \\
		\\
		$\frac{\del \bm{U}_n}{\del\bm{T}}\rvert_{T_\text{ref}} = 0.001\left( \frac{\splitfrac{0.005269056 + 3.299265709\theta_n - 91.79325798\theta_n^2 + 
				1004.911008\theta_n^3 - 5812.278127\theta_n^4}{ +
				19329.7549\theta_n^5 - 37147.8947\theta_n^6 + 38379.18127\theta_n^7 -
				16515.05308\theta_n^8}}{\splitfrac{1 - 48.09287227\theta_n + 1017.234804\theta_n^2 - 10481.80419\theta_n^3 +
				59431.3\theta_n^4 }{- 195881.6488\theta_n^5 + 374577.3152\theta_n^6 -385821.1607\theta_n^7 + 165705.8597\theta_n^8}} \right) $\\
			\\
			$\theta_p = \frac{c^*_{s,p}(x,t)}{c^{\text{max}}_{s,p}}$ \\
	$\theta_n = \frac{c^*_{s,n}(x,t)}{c^{\text{max}}_{n,p}}$ \\
	Open circuit potential \\
	\hline \\
	$U_{\text{p,ref}} =  \frac{-4.656 + 88.669\theta_p^2 - 401.119\theta_p^4 + 342.909\theta_p^6 -  462.471\theta_p^8 + 433.434\theta_p^10}{
	-1 + 18.933\theta_p^2 - 79.532\theta_p^4 + 37.311\theta_p^6 - 73.083\theta_p^8 + 95.96\theta_p^10}$\\
\\
	$U_{\text{n,ref}} =  0.7222 + 0.1387\theta_n + 0.029\theta_n^{0.5} - \frac{0.0172}{\theta_n}+ \frac{0.0019}{\theta^{1.5}} + 0.2808e^{0.9 - 15\theta_n} - 0.7984e^{0.4465\theta_n - 0.4108}$
	\\
	\\
	Heat source terms (electrodes) \\
	\hline \\
	$\bm{Q}_{\text{ohm}} = \sigma_{\eff, i}\left( \frac{\del \Psi(x,t)}{\del x} \right)^2 + \kappa_{\eff,i}\left(\frac{\del \Phi_e(x,t)}{\del x}\right)^2 + \frac{2\kappa_{\eff,i}RT(x,t)}{F}(1-t_+)\frac{\del \ln u(x,t)}{\del x} \frac{\Phi_e(x,t)}{\del x}$\\
	$\bm{Q}_{\text{rxn}} = Fa_ij(x,t)\eta_i(x,t)$\\
	$\bm{Q}_{\text{rev}} = Fa_ij(x,t)T(x,t)\frac{\del U_i}{\del T}\rvert_{T_\text{ref}}$\\
	\\
	Heat source terms (separator) \\ \hline \\
	$\bm{Q}_{\text{ohm}} =  \kappa_{\eff,i}\left(\frac{\del \Phi_e(x,t)}{\del x}\right)^2 + \frac{2\kappa_{\eff,i}RT(x,t)}{F}(1-t_+)\frac{\del \ln u(x,t)}{\del x} \frac{\Phi_e(x,t)}{\del x}$ \\
	\\
	Coefficients \\
	\hline \\
	\\
	$\bm{D}_{\eff,i} = \epsilon_i^{\text{brugg}_i} \times 10^{-4} \times 10^{-4.43 - \frac{54}{T(x,t)-229-5\times10^{-3}u(x,t)} - 0.22\times10^{-3}u(x,t)}$\\
	$\begin{aligned}
	\kappa_{\eff,i} =\epsilon_i^{\text{brugg}_i} \times 10^{-4}\times u(x,t)\left(-10.5 + 0.668\times10^{-3}u(x,t) + 0.494\times10^{-6}u(x,t)^2 + \right. \\
	\left. T(x,t)(0.074-1.78\times10^{-5}u(x,t) - 8.86\times10^{-10}u(x,t)^2 ) + \right. \\ \left. T(x,t)^2(-6.96\times10^{-5}+2.8\times10^{-8}u(x,t)) \right)^2
	\end{aligned}$\\
	$k_{\eff} = k_i e^{\frac{E^{k_i}_a}{R}\left(\frac{1}{T(x,t)} - \frac{1}{T_{\text{ref}}}\right)}$\\
	$D^s_{\eff,i} = D^s_i e^{\frac{E^{D^s_i}_a}{R}\left(\frac{1}{T(x,t)} - \frac{1}{T_{\text{ref}}}\right)} $\\
	$\sigma_{\eff, i} = \sigma_i(1-\epsilon_i-\epsilon_{f,i})$\\
        %\todo{cbm: is ``:='' intentional?}
	$\Upsilon = \frac{2(1-t_+)R}{F} $
	\end{longtable}
\end{center} 
\newpage
\section{Parameters used in simulation}\label{sec:appen_param}
\begin{center}
	\footnotesize
	\begin{tabular}{ ccccccc}
		 & &  
		Al CC & Cathode & Separator &
		Anode & Carbon CC \\
		\hline
		$c_e^{\text{init}}$ & [mol/m$^3$] & - & 1000 & 1000 & 1000 & - \\
		$c_s^{\text{avg, init}}$ & [mol/m$^3$] & - & 25751 & - & 26128 & - \\
		$c_s^{\text{max}}$ & [mol/m$^3$] & - & 51554 & - & 30555 & - \\
		$D_i$ & [m$^2/s$] & - & $7.5 \times 10^{-10}$ & $7.5 \times 10^{-10}$ & $7.5 \times 10^{-10}$ & -\\
		$D^s_i$ & [m$^2/s$] & - & $10^{-14}$ & -  & $3.9 \times 10^{-14}$ & -  \\
		$k_i$ & [m$^{2.5}/$ (mol$^{0.5}$s)] & - & $2.334\times 10^{-11} $& -  & $5.301 \times 10^{-11}$ & -\\
		$l_i$ & [m] & $10^{-5}$& $8\times 10^{-5}$ & $2.5 \times 10^{-5}$ & $8.8 \times 10^{-5}$ & $10^{-5}$ \\
		$R_{p,i}$ & [m] & - & $2\times 10^{-6}$ & - & $2\times 10^{-6}$ & -  \\
		$\rho_i$ & [kg/m$^3$]& 2700 & 2500 & 1100 & 2500 & 8940 \\
		$C_{p,i}$ & [J/(kg K)] & 897 & 700& 700& 700& 385 \\
		$\lambda_i$ & [W/(m K)] & 237& 2.1& 0.16& 1.7 & 401 \\
		$\sigma_i$ & [S/m] & $3.55 \times 10^{7}$ & 100 & - & 100 & $5.96 \times 10^{7}$ \\
		$\epsilon_i$ & - & - & 0.385 & 0.724 & 0.485 & - \\
		$a_i$ & [m$^2$/m$^3$]&  - & 885000 & - & 723600 & - \\
		$E^{D^s_i}_a$ & [J/mol] & -& 5000 & - & 5000 & - \\
		$E^{k_i}_a$ & [J/mol]&- &5000 &- &5000 &- \\
		brugg &- &- &4 & 4& 4 & - 
		\\
		$F$ & 96485 [C/mol] & -& -& -&- &-
		\\
		$R$ & 8.314472  [J/(mol K)]&- &- &- &- &- \\
		$t_{+}$ & 0.354 & & & & & \\
		$\epsilon_{f,i}$ & -& -&0.025 &- &0.0326 &-
		
	\end{tabular}
\end{center}


\bibliographystyle{elsarticle-num}

\bibliography{sample}

\end{document}